\documentclass{elsart}

\usepackage{harvard}

 \usepackage{graphics}
\usepackage{graphicx}
\usepackage{epsfig}

\usepackage{amssymb}

\def\url#1{{\ttfamily\def\/{/\discretionary{}{}{}}#1}}

\begin{document}

\begin{frontmatter}
\title{Dark matter distribution function from non-extensive statistical mechanics}
\author{Steen H. Hansen, Daniel Egli, }
\author{Lukas Hollenstein, Christoph Salzmann}
\address{University of Zurich, Winterthurerstrasse 190,
8057 Zurich, Switzerland}

\begin{abstract}
We present an analytical and numerical study of the velocity
distribution function of self gravitating collisionless particles,
which include dark matter and star clusters.  We show that the
velocity distribution derived through the Eddington's formula is
identical to the analytical one derived directly from the generalized
entropy of non-extensive statistical mechanics. This implies that self
gravitating collisionless structures are to be described by
non-extensive thermo-statistics. We identify a connection between the
density slope of dark matter structures, $\gamma$, from $\rho \sim
r^{-\gamma}$, and the entropic index, $q$, from the generalized
entropy, $S_q$.  
Our numerical result confirms the analytical findings of earlier
studies and clarifies which is the correct connection between the
density slope and the entropic index.
We use this result to conclude that from a
fundamental statistical mechanics point of view the central density
slope of self gravitating collisionless dark matter structures is not
constrained, and even cored dark matter structures are allowed with
$\gamma = 0$.
We find that the outer density slope is bounded by $\gamma= 10/3$.
\end{abstract}

\begin{keyword}
cosmology: dark matter --- gravitation ---
galaxies: kinematics and dynamics ---
cosmology: theory --- methods: analytical
\end{keyword}
\end{frontmatter}


\section{Introduction}
The existence of dark matter (DM) has been established beyond any doubt,
and the formation and evolution of cosmological structures through
gravitational attraction is reasonably well understood. The velocity
distribution of the resulting structures, however, remains unknown.
This is in stark contrast to the well established velocity
distribution of ideal gases, $f(v) \sim {\rm exp}(-mv^2/2kT)$, which
is derived from the extensive Boltzmann-Gibbs entropy. Dark matter,
however, experiences long-range gravitational interaction, and may
therefore not obey the rule of extensivity.

At the same time numerical simulations provide predictions of steep
central density cusps with power law slopes, $\rho \sim r^{-\gamma}$,
with $\gamma$ from $1$ to $1.5$ within a few percent of the virial
radius of the halo~\cite{nfw,moore}. However, recent careful
studies~\cite{diemand,reed,stoehr,navarro} indicate that the resolved region
has still not converged on a central density slope.  Analytical
analyses of the Jeans equations seem to indicate that the most shallow
allowed slope is $\gamma=1$~\cite{hansen}.  Such steep inner
numerically resolved slopes are, however, not supported by
observations.  By measuring the rotation curve of a galaxy one can in
principle determine the density profile of its DM halo. Low surface
brightness galaxies and spirals, where the observed dynamics should be
DM dominated, seem to show slowly rising rotation curves
\cite{rubin85,courteau97,palunas00,blok01,blok02,salucci01,swaters02,corb03}
indicating that these DM halos have constant density cores.  Galaxy
clusters, where baryons can play even less of a role, may show a
similar discrepancy. Arcs \cite{sand02} and strong lensing fits of
multiple image configurations and brightnesses \cite{tyson98} also
indicate shallow cores in clusters.

Facing this apparent disagreement between observations and numerical
simulations it is very important to understand if the pure dark matter
central density slopes are constrained from a fundamental statistical
mechanics point of view. We will here combine an analytical and a
numerical approach.  First we show that the dark matter distribution
functions are exactly the ones derived from non-extensive statistical
mechanics.  We then find a connection between the density slope,
$\gamma$, and the entropic index, $q$. A given theoretical bound on
$q$ will then lead to a bound on $\gamma$. We show that the bounds we
can imagine still allow any density slope, including a core with
$\gamma=0$.  The outer slope is bounded by $10/3$, which is close to
the findings of N-body codes.  Our findings can be very useful and
directly applicable if one can envisage stronger theoretical bounds on
$q$.

\section{Eddington's formula}
Let us first derive the actual velocity distribution function of
self gravitating collisionless structures numerically.

We will be considering the simplest possible dark matter structures
(actually any collisionless system, including star clusters), which
are spherical and isotropic. According to recent numerical N-body
results this is a reasonably good approximation in the central
equilibrated part of dark matter structures. We will argue later that
statistical mechanics in a natural way forces a system towards
isotropy.  Any given density profile, $\rho(r)$, can be
inverted~\cite{eddington} to give the particle velocity distribution
function, $f(E)$, through
\begin{equation}
f(E) = \frac{1}{\sqrt{8} \pi^2} \int_0 ^E 
\frac{d^2\rho}{d \Psi^2} \frac{d \Psi}{\sqrt{E-\Psi}} \, ,
\end{equation}
where $\Psi(r)$ is the relative potential as a function of radius, and
$E$ is the relative energy, $E =\Psi -mv^2/2$. For recent applications
and technical details see~\cite{binney,stelios}. For the numerical
inversion we typically consider structures with 10 orders of magnitude
in radius, and always confirm that the resulting velocity
distributions have converged. We consider only isotropic structures
and leave non-isotropy for a future analysis.  For simplicity we
construct very simple structures which depend only on one free
parameter, namely the density slope $\gamma$, from $\rho (r) \sim
r^{-\gamma}$. The resulting velocity distribution functions are
presented as (coloured) symbols in figure 1 for the slopes
$\gamma=1.0, 1.5, 2$ and $2.3$.

\begin{figure}[b]
\begin{center}
\includegraphics[height=6.5cm,width=8cm,angle=0]{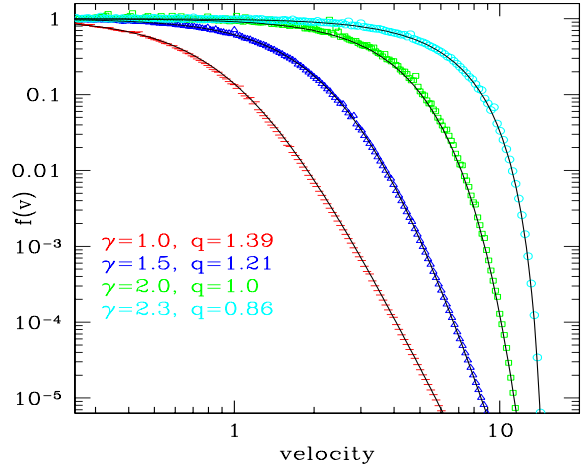}
\end{center}
\caption{The velocity distribution functions. The (coloured) symbols
are from inverting a given density distribution according to
Eddington's formula, for the density slopes of $\gamma = 1.0, 1.5,
2.0$ and $2.3$. The solid lines are the theoretical formulae from
Tsallis statistics, using entropic indices of $q=1.39, 1.21, 1.0$ and
$0.86$. The curves are arbitrarily normalized, and use different velocity
dispersion, $\sigma^2$, in order to make the figure more readable and
the difference in shapes more evident. The labels describe the lines 
from left to right.}
\label{fig:vel.distr}
\end{figure}

It is important to note that for $\gamma >2$ there is a maximal
velocity.  This is a real feature, purely related to the density
profile, and is not simply the result of a finite structure. This
shows directly that the velocity distribution function cannot be an
exponential or any sum of exponentials, since exponential functions
have tails going to infinite energy. We have thus shown explicitely
that the often discussed sum of
exponentials~\cite{lyndenbell,kull,iguchi} cannot be correct, and
instead a more general functional form for the velocity distribution
function must be sought. We will now turn to non-extensive
thermo-statistics and show that the correct functional form for the
velocity distribution function is easy to derive.

\section{Tsallis statistics}
Statistical mechanics for classical gases can be derived from the
Boltzmann-Gibbs assumption for the entropy, $S_{BG} = - k \sum p_i \cdot 
{\rm ln}p_i$,
where $p_i$ is the probability for a given particle to be in the state
$i$, and the sum is over all states. For normal gases the probability,
$p(v)$,
coincides with the velocity distribution function, $f(v)$.
This classical statistics can be 
generalized to Tsallis (or non-extensive) statistics~\cite{tsallis88}, 
which depends on 
the entropic index $q$
\begin{equation}
S_q = - k \sum_i p_i^q \cdot {\rm ln}_q p_i \, ,
\label{stsallis}
\end{equation}
where the q-logarithm is defined by, ln$_q p =
(p^{1-q}-1)/(1-q)$, and for $q=1$ the normal Boltzmann-Gibbs entropy
is recovered, $S_{BG} = S_1$.  The probabilities still obey, $\sum
p_i =1$, while the particle distribution function is now given by
$f(v) = p^q(v)$. Thus, for $q<1$ one privileges rare events, whereas
$q>1$ privileges common events.  
One also sees directly that entropy is maximized for isotropic
distributions, which supports our original isotropy assumption.
For a summary of applications
see~\cite{tsallis99}, and for up
to date list of references see  {\tt
http://tsallis.cat.cbpf.br/biblio.htm}.

Average values are calculated through the particle distribution function,
and one e.g. has the mean energy~\cite{tsallisMP98}
\begin{equation}
U_q = \frac{\sum p_i^q E_i}{c_p} \, ,
\end{equation}
where $c_p=\sum p_i^q$, and $E_i$ are the energy eigenvalues. 
Optimization of the entropy in eq.~(\ref{stsallis})
under the constraints leads to
the probability~\cite{silva98,tsallis03}
\begin{equation}
p_i = \frac{\left[1 - (1-q)\beta_q \left(E_i-U_q\right)  
\right] ^{1/(1-q)}}{Z_q} \, ,
\label{eq:pi1}
\end{equation}
where $Z_q$ normalizes the probabilities, $\beta_q = \beta/c_p$, and
$\beta$ is the optimization Lagrange multiplier associated with the
average energy.  Adding a constant energy, $\epsilon_0$, to all the
energy eigenvalues leads to $U_q \rightarrow U_q + \epsilon_0$, which
leaves all the probabilities, $p_i$, invariant~\cite{tsallisMP98}.
When $\alpha$ is positive, where
\begin{equation}
\alpha = 1 +  (1-q)\beta_q U_q \, ,
\label{eq:alpha}
\end{equation}
then eq.~(\ref{eq:pi1}) can be written as
\begin{equation}
p_i = \frac{
\left( 1 - (1-q) (\beta_q/\alpha) \,  E_i
\right) ^{1/(1-q)}}{Z_q'} \, .
\label{eq:pi2}
\end{equation}
On figure 1 we plot 4 solid (black) lines, corresponding to $q=1.39,
1.21, 1.0$ and $0.86$, and well chosen $\beta$'s.

It is clear from figure 1 that the velocity distribution function,
$f(v)$, inverted from a density distribution function with slope
$\gamma =2.3$, where $\rho \sim r^{-\gamma}$, is extremely well fitted
with the non-extensive form in~eq.~(\ref{eq:pi2}) using
$q=0.86$. Similarly, the Maxwell distribution (the exponential, $q=1$)
is recovered with an isothermal sphere (with $\gamma=2$), and finally,
the $\gamma =1.5$ is well fitted with $q=1.21$. One should keep in
mind that the probabilities, $p(v)$, are related to the physical
distributions through $f(v)=p^q(v)$.

We consider this excellent agreement in figure 1 as a strong indicator that
equilibrium self gravitating structures indeed are to be described by
non-extensive statistical mechanics. We are thus giving theoretical
support to the use of Tsallis entropy (and the resulting velocity
distribution function) when analysing gravitational structures. Such
analysis was first conducted in~\cite{lavagno}, where an entropic
index of $q\approx 0.23$ was found to describe accurately the peculiar
velocity of galaxy clusters.  It is worth emphasizing that the maximal
velocity for $q<1$, $v_{\rm max} = \sqrt{2 \alpha /m\beta_q(1-q)}$ is
an inherent property of the non-extensive thermo-statistics (for
$q<1$) and is not related to any finite size of the system. We also
note that the $\alpha$ in eq.~(\ref{eq:alpha}) may be important for 
cluster temperature observations~\cite{clustertemp}.

The structures we are considering are polytropes, which are described
by the polytropic index, $n$, making the connection between pressure
and density, $p \sim \rho^{1+1/n}$. It was first shown analytically by
\cite{plastino93} that these structures are
exactly described by non-extensive statistical mechanics, and a
connection was found between $n$ and the entropic index, $q$.  Thus,
the finding of this section is a numerical confirmation of the
original work of \cite{plastino93}, which
says that the distribution function for polytropes is infact given by
eq.~(\ref{eq:pi2}).  As we will discuss below, the actual connection
between $n$ and $q$ is not clear from the literature, and our
independent results are useful in distinguishing which of the
analytical results is correct.

Naturally, for a more general density profile where $\gamma$ is not
constant, the actual velocity distribution function will be a sum over
terms of the shape given by eq.~(\ref{eq:pi2}), with $f(v) = p^q(v)$.  
One can only
approximate the velocity distribution function with one term of the
form given in eq.~(\ref{eq:pi2}) when $\gamma(r)$ varies sufficiently
slowly with radius.

\section{A connection between the density slope and the entropic index?}

We can now proceed as in the previous section with various density
profiles, and find the corresponding $q$. The result is presented in
figure 2. The fitted $q$'s are accurate at
the few percent level.
This figure is presented only in the range $0< \gamma <
2.5$, since steeper slopes lead to numerical problems in our simple
inversion of Eddington's formula.  This is not a problem
since we are interested in exactly this range in density profiles. 
As pointed out in section 3, our numerical finding that the
self-gravitating structures are described by non-extensive statistical
mechanics was first shown analytically by 
\cite{plastino93}.
There seems, however, to be some confusion in the literature, as to
what the connection between $n$ and $q$ really is. Polytropes follow
the Lane-Emden equation, and for power-law density profiles one thus has,
$n=\gamma/(\gamma-2)$. Now, there are 3 different connections between
$n$ and $q$ frequently used in the literature (see e.g. 
\cite{taruya03,chavanis,tsallis03} for different expressions) and
we can therefore use our independent numerical method to distinguish which
is the correct one. We find that the formula derived 
in \cite{tsallis03,limasouza,silvaAlcaniz}, 
$n = (5-3q)/(2(1-q))$ fits our results very
well. We can therefore conclude that the correct connection between
the density slope, $\gamma$, and the entropic index, $q$, is given by
\begin{equation} 
q = \frac{10-3\gamma}{6-\gamma} \, .
\label{eq:qgamma}
\end{equation} 
This is shown as the thin (green) solid line in figure 2.
We note that this equation may play an important
role in explaining the observed correlation between the density slope
and the anisotropy of dark matter structures~\cite{abrelation}.

\begin{figure}[b]
\begin{center}
\includegraphics[height=6.5cm,width=8cm,angle=0]{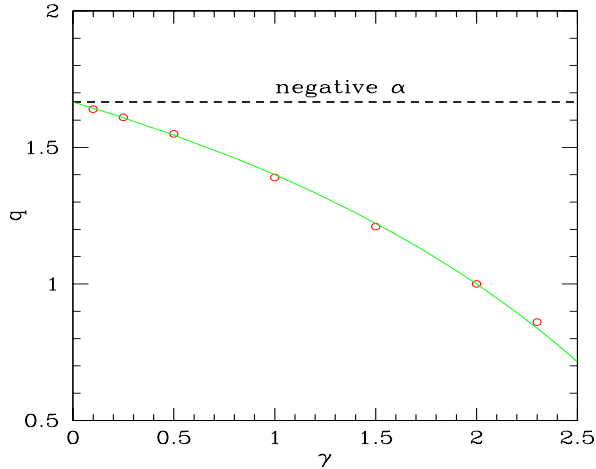}
\end{center}
\caption{A given density profile, $\rho \sim r^{-\gamma}$, gives a
specific distribution function, $f(v)$, which can be fitted with two
free parameters, namely the entropic index, $q$, and the velocity
dispersion, $\sigma^2$. We present the resulting $q$ as a function of
the density slope, $\gamma$. The theoretical limit
of $q<5/3$ is presented with a dashed line. The thin (green) solid 
line is the curve given in eq.~(\ref{eq:qgamma}).}
\label{fig:beta.q}
\end{figure}

It is interesting to ask the question: 
which are the theoretical constraints on the
entropic index $q$?  We saw above that eq.~(\ref{eq:pi2}) is only
valid for positive $\alpha$.  Actually the expression in
eq.~(\ref{eq:pi2}) is also valid for negative $\alpha$, however, a
negative effective temperature (a negative velocity dispersion in our
case), $\beta_q' = \beta_q/\alpha$ is difficult to interpret for
cosmological structures.  Since we find numerically that $U_q\beta_q =
3/2$, one sees that $\alpha=0$ for $q=5/3$.  The same conclusion,
namely $q<5/3$, was reached by \cite{boghosian} by
demanding a positive proportion between temperature and internal
energy; and also reached by \cite{silvaAlcaniz}
through an analysis of the velocity dispersion.

We see in figure 2, that $q=5/3$ corresponds to a cored dark matter
distribution with $\gamma =0$. Thus all dark matter density slopes are
allowed from a fundamental statistical mechanics point of view.  Our
findings are directly applicable if one can imagine a stronger
theoretical bound on $q$. E.g. \cite{boghosian}
considered the ideal gas case, and found that positiveness of the
thermal conductivity leads to \mbox{ $q<7/5$}. Such a bound would
according to our figure 2 imply a bound on the central density profile
of $\gamma >1$, in agreement with numerical and analytical
suggestions~\cite{diemand,reed,navarro,hansen}. It has indeed been
suggested~\cite{weike} that the numerical methods induce an artificial
conductivity, which would then imply a central slope of $-1$.

The lower bound on $q$ is given by $q>0$~\cite{lima02}, which implies
a bound on the dark matter profile of $\gamma < 10/3$, when using
eq.~(\ref{eq:qgamma}).  This is very close to the findings of
numerical simulations, which is $\gamma \leq 3$. Further, if one would
have a bound of $q \geq 1/3$ (as suggested by~\cite{lima01}), then one
would have a bound on the outer density profile of $\gamma \leq 3$.

\section{Conclusions}
We compare the velocity distribution function for self gravitating
collisionless structures (e.g. dark matter or star clusters) with the
analytical expectation from non-extensive statistical mechanics, and
find excellent agreement between the two.  This is a strong indication
that self gravitating systems of collisionless particles must be
described by non-extensive statistical mechanics.

We identify a connection between the density power slope, $\gamma$,
from $\rho \sim r^{-\gamma}$, and the entropic index, $q$, from the
generalized entropy, $S_q$. We show that $q$ is only bounded from
above by $q<5/3$, which implies that all dark matter central slopes
are allowed from this fundamental statistical mechanics point of view,
including central cores with $\gamma=0$. We show that the outer
density slope is bounded by $\gamma \leq 10/3$.  If one can imagine
stronger theoretical bounds on $q$ then our findings are directly
applicable to give stronger bounds on the allowed density slope.

\section*{Acknowledgements}
It is a pleasure to thank Marcel Zemp for discussions, and
Sara Hansen for inspiration.
SHH thanks the Tomalla
foundation for financial support.


\end{document}